\documentclass[conference]{IEEEtran}
\IEEEoverridecommandlockouts

\usepackage{
amsmath,
graphicx,
cite,
multirow,
booktabs,
amssymb,
enumitem,
url,
amsfonts,
algorithmic,
textcomp,
xcolor,
hyperref,
tabularx
}

\def\BibTeX{{\rm B\kern-.05em{\sc i\kern-.025em b}\kern-.08em
    T\kern-.1667em\lower.7ex\hbox{E}\kern-.125emX}}
    
\begin{document}

\title{DRASP: A Dual-Resolution Attentive Statistics Pooling Framework for Automatic MOS Prediction}

\author{\IEEEauthorblockN{
Cheng-Yeh Yang\IEEEauthorrefmark{1},
Kuan-Tang Huang\IEEEauthorrefmark{1},
Chien-Chun Wang\IEEEauthorrefmark{1},
Hung-Shin Lee\IEEEauthorrefmark{3},
Hsin-Min Wang\IEEEauthorrefmark{2},
and Berlin Chen\IEEEauthorrefmark{1}
}
\IEEEauthorblockA{\IEEEauthorrefmark{1}Dept. Computer Science and Information Engineering, National Taiwan Normal University, Taiwan}
\IEEEauthorblockA{\IEEEauthorrefmark{2}Institute of Computer Science, Academia Sinica, Taiwan}
\IEEEauthorblockA{\IEEEauthorrefmark{3}United Link Co., Ltd., Taiwan}
}

\maketitle

\begin{abstract}

A pooling mechanism is essential for mean opinion score (MOS) prediction, facilitating the transformation of variable-length audio features into a concise fixed-size representation that effectively encodes speech quality.
Existing pooling methods typically operate at a singular granularity, concentrating either on a comprehensive global perspective or a detailed frame-level analysis, which may overlook complementary perceptual insights.
To address this limitation, we introduce the Dual-Resolution Attentive Statistics Pooling (DRASP) framework.
DRASP integrates both coarse-grained, global statistical summaries and fine-grained, attentive analyses of perceptually significant segments.
This dual-view architecture empowers our model to formulate a more thorough and robust representation, capturing both the overarching structural context and salient local details concurrently.
Extensive experiments validate the effectiveness and strong generalization ability of the proposed framework.
It consistently outperforms various baseline methods across diverse datasets (MusicEval and AES-Natural), MOS prediction backbones (including a CLAP-based model and AudioBox-Aesthetics), and different audio generation systems, achieving a relative improvement of 10.39\% in system-level Spearman's rank correlation coefficient (SRCC) over the widely-used average pooling approach.

\end{abstract}


\section{Introduction}

The field of generative audio has witnessed a recent surge, giving rise to sophisticated systems for text-to-music (TTM), text-to-speech (TTS), and text-to-audio (TTA) applications \cite{chen2024a,ziv2024,chen2024,chen2025,majumder2024}.
As these systems generate increasingly complex and high-fidelity audio, the demand for reliable and scalable evaluation methodologies has become critical.
Traditional evaluation approaches, however, encounter significant challenges—objective metrics, such as the widely adopted Fréchet Audio Distance (FAD) \cite{kilgour2019}, frequently fail to correlate with human auditory perception, while subjective evaluations, including mean opinion score (MOS) studies, are resource-intensive and challenging to scale.
To overcome this evaluation bottleneck, the development of models for MOS prediction has emerged as a promising research direction.

A key component in MOS prediction architectures is the temporal pooling layer, which aggregates frame-level features into a fixed-size vector.
Among conventional approaches, statistics pooling \cite{snyder2017} represents a significant advancement over simple average pooling.
By computing not only the mean but also the standard deviation, it captures the temporal variation across the entire sequence, thus providing a comprehensive, global summary of the dynamic characteristics present in the audio signal.
Despite this advancement, the static nature of statistics pooling poses a fundamental limitation.
The indiscriminate averaging across all frames can lead to a skewed global perspective, as perceptually critical local events---such as artifacts or highlights---are often diluted by the presence of longer, less informative segments.

To address this, attentive statistics pooling \cite{okabe2018} was introduced to identify and focus on these perceptually important local regions.
By incorporating an attention mechanism \cite{vaswani2017}, the model learns to assign varying weights to individual frames, enabling it to compute an importance-weighted mean and standard deviation.
This technique has been widely adopted and demonstrated efficacy in the speaker verification domain \cite{desplanques2022,wang2023,shon2019,jung2020}.
However, when applying this technique to MOS prediction, we observe that its strictly frame-level, or highly local, focus presents a critical limitation. A purely local attention mechanism can be excessively sensitive to transient noise while neglecting the broader global acoustic patterns vital for comprehensive quality perception.
This observation is consistent with our own experiments, where attentive statistics pooling did not consistently outperform the simpler global statistics pooling---an issue also noted in previous studies \cite{deng2023}.

Building on these insights, we propose DRASP, a \textbf{D}ual-\textbf{R}esolution \textbf{A}ttentive \textbf{S}tatistics \textbf{P}ooling framework, designed to leverage the complementary strengths of both global and locally-focused pooling.
The architecture of DRASP is founded on a dual-branch design.
The first component is a global statistics branch that, aligning with the advantages of statistics pooling, computes the overall mean and standard deviation across the entire utterance to capture its holistic temporal dynamics.
The second component is a fine-grained attentive branch.
It inherits the core idea of attentive statistics pooling by using an attention mechanism to focus on salient information.
Crucially, we advance this concept by operating at the segment level \cite{liu2019}.
Rather than weighing individual frames, this branch learns to identify and prioritize the importance of meaningful acoustic segments, thereby enhancing robustness to local noise.
Moreover, a trainable fusion mechanism adaptively combines the outputs of these two branches into a single and powerful representation.

Our main contributions are as follows:
\begin{enumerate}[label=\arabic*),leftmargin=14pt,itemsep=0ex,topsep=4pt]
\item \textbf{Dual-Resolution Framework:} We propose DRASP, a framework that combines global statistical features with a local, fine-grained attentive summaries to overcome the limitations of single-granularity pooling strategies.
\item \textbf{Segmental Attention for Robust MOS Prediction:} To the best of our knowledge, we are the first to adapt a segmental attention mechanism to specifically address the instability of frame-level attentive statistics pooling in the context of MOS prediction.
By shifting the attentional focus from noisy individual frames to more meaningful acoustic segments, our framework achieves more robust and perceptually relevant quality predictions.
\item \textbf{Comprehensive Empirical Validation:} We demonstrate through extensive experiments that DRASP achieves superior performance and generalizability, consistently outperforming baselines across diverse datasets (MusicEval \cite{liu2025} and AES-Natural \cite{tjandra2025}), audio generation systems (TTS, TTM, and TTA), and MOS prediction backbones (the CLAP-based model \cite{wu2023} and AudioBox-Aesthetics \cite{tjandra2025}).
\end{enumerate}

\section{Related Work}

\subsection{From Static to Adaptive Pooling Mechanisms}

While conventional methods such as statistical pooling \cite{snyder2017} capture temporal variations by computing the mean and standard deviation, their inherent static nature---treating all frames equally---poses a significant limitation.
To address this, attention mechanisms have been introduced to dynamically weight perceptually important frames \cite{cai2018}. This has led to the development of attentive statistics pooling \cite{okabe2018}, which computes an importance-focused mean and standard deviation.
This concept has been further refined through various approaches, such as leveraging multi-head attention to capture different audio aspects \cite{zhu2018,india2019}, employing vector-based attention for finer-grained control \cite{wu2020}, and developing more robust mechanisms like SSDP \cite{deng2023}.
These studies collectively demonstrate a clear research trajectory toward creating more adaptive pooling strategies to better model temporal importance.

\subsection{Multi-Scale and Multi-Resolution Approaches}

An important research direction for enhancing pooling layers involves integrating information across diverse temporal granularities, typically achieved through multi-scale and multi-resolution strategies.

Multi-scale approaches generally aggregate features from various layers of the network backbone.
A representative method is MTSP \cite{wu2022}, which extracts and pools feature maps from multiple distinct stages of a convolutional neural network to construct a rich and hierarchical representation.
While this approach is effective, its definition of scale is constrained by the architectural depth of the feature extractor.
In contrast, our DRASP framework generates multiple views from a single input layer, where each view corresponds to a different temporal analysis scope, such as global or segmental.
This design enables a more direct modeling of temporal quality variations that closely align with human judgments.

The concept of multi-resolution has also been integrated into attention mechanisms.
A relevant approach \cite{wang2020} generates multiple levels of resolution by employing several attention heads, each configured with a distinct temperature parameter in the softmax function.
This configuration enables certain heads to concentrate narrowly while allowing others to adopt a broader perspective, effectively defining resolution through \textit{the sharpness of attention weights}.
In contrast, DRASP defines resolution through \textit{the structure of the input} itself.
Specifically, it analyzes either the entire utterance or smaller discrete segments, thereby creating an explicit distinction in granularity.
This structural formulation offers a clearer and more consistent integration of both holistic and localized information, as opposed to relying on the implicit differentiation learned by attention heads.

\begin{figure}[t]
\centering
\includegraphics[width=1.0\linewidth]{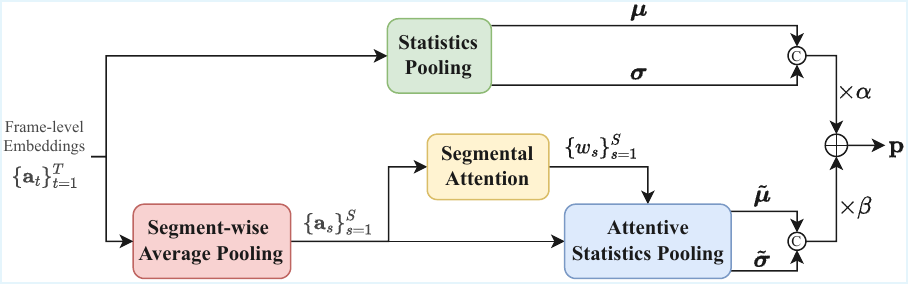}
\vspace{-15pt}
\caption{Overview of the proposed DRASP mechanism.}
\label{fig:main}
\vspace{-15pt}
\end{figure}

\subsection{Adaptive and Quality-Aware Pooling Methods}

The challenge of adaptively aggregating features from inputs with varying levels of quality extends beyond audio processing.
This issue has been extensively investigated within the domain of computer vision, particularly in the context of set-based face recognition.
A notable advancement in this domain is NetVLAD \cite{arandjelovic2016}, which introduces a learnable and end-to-end pooling layer that surpasses conventional averaging strategies.
To effectively address low-quality images, this method was further enhanced through GhostVLAD \cite{zhong2018}, which incorporates ``ghost clusters'' as a form of null space.
This design enables the model to eliminate uninformative features during the aggregation process.

The concept of quality-aware pooling has been explored through various innovative approaches.
For instance, NAN \cite{yang2017} employs an explicit attention mechanism to direct the pooling process, whereas QAN \cite{liu2017} predicts a quality score for each input to minimize the impact of low-quality samples.
These methods, originating from distinct domains, underscore the broader significance of developing sophisticated pooling strategies that can selectively filter or weight inputs in accordance with their quality.

\section{Proposed Method}

To address the limitations of traditional pooling methods in capturing perceptually salient audio patterns, we introduce the Dual-Resolution Attentive Statistics Pooling (DRASP) mechanism.
Our framework is founded on the principles of attentive statistics pooling \cite{okabe2018} and deep segment attentive embedding \cite{liu2019}, seamlessly integrating them into a novel \textbf{dual-branch} architecture that concurrently processes audio features at two distinct granularities.
As shown in Fig. \ref{fig:main}, one branch captures a coarse-grained, global overview, while the other performs a fine-grained, attentive analysis on audio segments.

\subsection{Global Overview Branch}

Let the input audio be represented as a sequence of frame-level embeddings $\{\mathbf{a}_t\}_{t=1}^T$, where each $\mathbf{a}_t \in \mathbb{R}^d$.
The global branch summarizes the entire utterance by computing basic, unweighted statistics across all frames.
A standard \textit{statistics pooling} layer is applied to the full sequence $\{\mathbf{a}_t\}_{t=1}^T$ to obtain the global mean vector $\boldsymbol{\mu}$ and standard deviation vector $\boldsymbol{\sigma}$.
This branch provides a coarse but comprehensive representation of the temporal structure in the audio signal, capturing overall characteristics while avoiding an emphasis on local variations.

\subsection{Segmental Attention Branch}

Complementing the global branch, this branch focuses on perceptually salient regions of the audio by operating at the segment level.
This branch implements our proposed \textbf{segmental attentive statistics pooling} architecture, which enhances local sensitivity while ensuring statistical robustness.

The frame-level sequence is initially partitioned into $S = T / n$ non-overlapping segments.
For each segment, a segment-level embedding $\mathbf{a}_s$ is computed using the \textit{segment-wise average pooling} module by averaging the $n$ corresponding frame embeddings:
\begin{equation}
\mathbf{a}_s = \frac{1}{n} \sum_{t=(s-1)n+1}^{sn} \mathbf{a}_t.
\end{equation}

Subsequently, a learnable \textit{segmental attention} mechanism assigns an importance weight $w_s$ to each segment embedding $\mathbf{a}_s$.
The attention score $z_s$ is computed using a trainable scoring function, followed by softmax normalization:
\begin{equation}
z_s = \mathbf{v}^\top \tanh(\mathbf{W} \mathbf{a}_s + \mathbf{b}),
\end{equation}
\begin{equation}
w_s = \frac{\exp(z_s)}{\sum_{i=1}^S\exp(z_i)},
\end{equation}
where $\mathbf{W}$, $\mathbf{b}$, and $\mathbf{v}$ represent the trainable parameters.

Using these weights, an \textit{attentive statistics pooling} layer computes the weighted mean $\tilde{\boldsymbol{\mu}}$ and the weighted standard deviation $\tilde{\boldsymbol{\sigma}}$:
\begin{equation}
\tilde{\boldsymbol{\mu}} = \sum_{s=1}^S w_s \mathbf{a}_s,~\tilde{\boldsymbol{\sigma}} = \sqrt{ \sum_{s=1}^S w_s \mathbf{a}_s \odot \mathbf{a}_s - \tilde{\boldsymbol{\mu}} \odot \tilde{\boldsymbol{\mu}} },
\end{equation}
where $\odot$ denotes element-wise multiplication.

By concentrating on short, acoustically significant segments and applying selective weighting, this approach effectively captures fine-grained temporal variations that are frequently overlooked by global pooling.
When combined with the global and coarse-grained branch, it provides a complementary representation that enhances the perceptual accuracy of the model.

\subsection{Adaptive Fusion of Global and Segmental Features}

To effectively leverage the complementary strengths of both the global and fine-grained branches, their outputs are integrated using a learnable linear fusion mechanism.
The final utterance-level representation $\mathbf{p}$ is computed as:
\begin{equation}
\mathbf{p} = \alpha [\boldsymbol{\mu}; \boldsymbol{\sigma}] + \beta [\tilde{\boldsymbol{\mu}}; \tilde{\boldsymbol{\sigma}}],
\end{equation}
where $[\cdot;\cdot]$ denotes vector concatenation.
The scalar weights $\alpha$ and $\beta$ are trainable parameters, initialized to $1$ and $0$, respectively.
This initialization biases the model to initially rely on the stable, coarse-grained global statistics, which provide a robust baseline representation.
As training progresses, the model gradually learns to integrate the fine-grained, attention-weighted features from the local branch, enabling it to adaptively emphasize perceptually salient segments that enhance accuracy.
By jointly optimizing these fusion weights, the model achieves a balance between global context and local detail, resulting in a more expressive and accurate representation of audio quality.

\begin{figure}[t]
\centering
\includegraphics[width=0.95\linewidth]{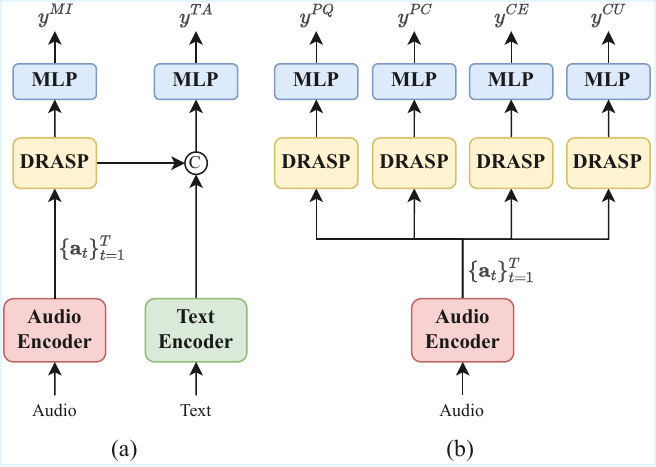}
\vspace{-10pt}
\caption{Architecture of two distinct MOS prediction backbones, each enhanced with the proposed DRASP module.}
\label{fig:backbones}
\vspace{-15pt}
\end{figure}

\begin{table*}[htbp]
\centering
\setlength{\tabcolsep}{1.5pt}
\caption{Performance comparison of various pooling methods on the MusicEval dataset.}
\vspace{-5pt}
\label{tab:track1}
\begin{tabular}{lcccccccccccc}
\toprule
\multirow{2}{*}{\bf{Method}} & \multicolumn{4}{c}{\bf{Musical Impression}} & \multicolumn{4}{c}{\bf{Textual Alignment}} & \multicolumn{4}{c}{\bf{Average}} \\
\cmidrule(r){2-5} \cmidrule(r){6-9} \cmidrule(r){10-13}
 & \bf{MSE $\downarrow$} & \bf{LCC $\uparrow$} & \bf{SRCC $\uparrow$} & \bf{KTAU $\uparrow$} & \bf{MSE $\downarrow$} & \bf{LCC $\uparrow$} & \bf{SRCC $\uparrow$} & \bf{KTAU $\uparrow$} & \bf{MSE $\downarrow$} & \bf{LCC $\uparrow$} & \bf{SRCC $\uparrow$} & \bf{KTAU $\uparrow$} \\
\toprule
Average Pooling & 0.299 & 0.856 & 0.838 & 0.628 & 0.125 & 0.862 & 0.836 & 0.648 & 0.212 & 0.859 & 0.837 & 0.638 \\
Statistics Pooling & \underline{0.102} & \underline{0.914} & 0.908 & 0.756 & \underline{0.060} & \underline{0.883} & \underline{0.885} & \underline{0.721} & \underline{0.081} & \underline{0.899} & 0.897 & 0.739 \\
Attentive Pooling & 0.179 & 0.878 & 0.895 & 0.738 & 0.098 & 0.835 & 0.838 & 0.652 & 0.139 & 0.857 & 0.867 & 0.695 \\
Attentive Statistics Pooling & 0.197 & 0.896 & 0.903 & 0.766 & 0.121 & 0.874 & 0.836 & 0.675 & 0.159 & 0.885 & 0.870 & 0.721 \\
Segmental Attentive Statistics Pooling & 0.157 & 0.903 & \underline{0.924} & \underline{0.798} & 0.099 & 0.869 & 0.877 & 0.712 & 0.128 & 0.886 & \underline{0.901} & \underline{0.755} \\
\midrule
\bf{DRASP} & \bf{0.076} & \bf{0.949} & \bf{0.957} & \bf{0.858} & \bf{0.058} & \bf{0.897} & \bf{0.890} & \bf{0.726} & \bf{0.067} & \bf{0.923} & \bf{0.924} & \bf{0.792} \\
\midrule
Multi-head Attentive Pooling & 0.187 & 0.889 & 0.912 & 0.766 & 0.102 & 0.875 & 0.849 & 0.675 & 0.145 & 0.882 & 0.881 & 0.721 \\
Multi-resolution Multi-head Attentive Pooling & 0.124 & 0.924 & 0.930 & 0.812 & 0.098 & 0.891 & 0.866 & 0.707 & 0.111 & 0.908 & 0.898 & 0.760 \\
\bottomrule
\end{tabular}
\vspace{-15pt}
\end{table*}

\subsection{Modular Integration with MOS Prediction Backbones}

To validate the versatility and effectiveness of DRASP as a modular component, we integrate it into two distinct MOS prediction backbones.
The first backbone, shown in Fig. \ref{fig:backbones} (a), is a multimodal framework for TTM evaluation.
We follow the design proposed in \cite{liu2025}, which utilizes both audio and text encoders from a pre-trained CLAP model \cite{wu2023}.
In this setup, DRASP replaces the original pooling layer of the CLAP audio encoder to generate a representative audio embedding $\mathbf{p}$.
The framework then employs two separate prediction heads: one multi-layer perceptron (MLP) processes the audio embedding $\mathbf{p}$ alone to predict the MOS score of musical impression $y^{MI}$, while another processes the concatenation of the audio embedding and the text embedding to predict the textual alignment score $y^{TA}$.

The second backbone, illustrated in Fig. \ref{fig:backbones} (b), enhances AudioBox-Aesthetics \cite{tjandra2025} by substituting its internal pooling mechanism with our proposed DRASP module.
In this setup, DRASP processes the frame-level features extracted from the original audio encoder to generate the utterance-level embedding $\mathbf{p}$.
This embedding is subsequently inputted into four parallel MLP heads, each tasked with predicting the corresponding perceptual dimensions: production quality ($y^{PQ}$), production complexity ($y^{PC}$), content enjoyment ($y^{CE}$), and content usefulness ($y^{CU}$).

\section{Experimental Setups}

\subsection{Datasets}

We evaluated our framework using two benchmark datasets, targeting both domain-specific and general-purpose audio quality prediction.
For the MOS prediction task of TTM models, we used the MusicEval dataset \cite{liu2025}, which contains 2,748 music clips generated by 31 distinct text-to-music systems.
Each system was conditioned on one of 384 unique text prompts.
Every clip received ratings from five annotators selected from a pool of 14 expert musicians, along two perceptual dimensions: musical impression and textual alignment.

To further evaluate the generalizability of our framework across diverse audio types, we employed the AES-Natural dataset\footnote{\url{https://github.com/facebookresearch/audiobox-aesthetics/tree/main/audiomos2025_track2}.}.
Due to the unavailability of some audio files from their original sources at the time of our experiments, we utilized a subset of the full dataset.
Our working subset comprises a total of 2,776 audio clips, encompassing speech, music, and general audio domains.
Specifically, it includes 950 speech samples sourced from EARS, LibriTTS, and Common Voice 13.0, 932 music samples drawn from MUSDB18-HQ and MusicCaps, and 894 general audio clips obtained from AudioSet.
Each audio sample was rated by 10 experts with professional backgrounds in audio or music.
All samples in AES-Natural were annotated along four perceptual dimensions: production quality (PQ), production complexity (PC), content enjoyment (CE), and content usefulness (CU).

\subsection{Configurations}

To validate the versatility of DRASP\footnote{Our open-source code: \url{https://github.com/jerryyang1231/DRASP/}.}, we conducted experiments using two distinct MOS prediction backbones.
For the setup involving the pre-trained CLAP \cite{wu2023} as the upstream feature extractor, we trained the model with a batch size of 64.
Optimization was performed using SGD with a fixed learning rate of 0.0005, minimizing the mean absolute error (MAE) as the loss function.
In contrast, when integrating DRASP with the pre-trained AudioBox-Aesthetics \cite{tjandra2025}, we adopted an alternative training configuration informed by the original paper.
Specifically, we used a batch size of 32 and the AdamW optimizer with a learning rate of 0.0001.
The model was trained using a composite loss function combining MAE and mean squared error (MSE).
For both configurations, early stopping was implemented based on the validation loss, with a patience of 20 epochs, to enhance training efficiency and mitigate the risk of overfitting.

\subsection{Evaluation Metrics}

We employed several standard metrics to objectively assess system performance, including MSE, linear correlation coefficient (LCC), Spearman's rank correlation coefficient (SRCC), and Kendall Tau rank correlation (KTAU), following the evaluation protocol described in \cite{cooper2022}.
These metrics were calculated by averaging MOS predictions across all generated clips for each system, and then comparing these aggregated predictions against corresponding human-annotated MOS scores at the system level.
Evaluating at the system level is crucial in practical applications, as it provides a more relevant assessment of overall speech system performance compared to individual clip evaluations.
MSE ranges from 0 to $\infty$, with lower values indicating better performance.
LCC, SRCC, and KTAU range from $-1$ to $1$, where higher values indicate stronger correlation and better alignment with human judgments.
Among these, SRCC is our primary metric of interest, as our evaluation focuses on system-level performance, where the main objective is to accurately rank different generative models in accordance with human perception.

\section{Results and Discussion}

\subsection{Results of Various Pooling Methods on MusicEval}

Table \ref{tab:track1} presents a comparison between our DRASP framework and several conventional pooling methods on the MusicEval dataset.
The results clearly demonstrate the effectiveness of the dual-branch architecture.
DRASP consistently outperforms all baselines in both perceptual dimensions, namely musical impression and textual alignment, achieving the lowest MSE and the highest correlations with human judgments as measured by LCC, SRCC, and KTAU.
These findings indicate that the integration of global and local temporal information through a dual-resolution mechanism yields a more robust and accurate representation of perceptual audio quality.

A closer examination of the baseline results reveals two key insights that informed the design of DRASP.
First, a clear performance hierarchy in SRCC reveals that while segmental attentive statistics pooling performs well, the frame-level variant \cite{okabe2018} underperforms even the non-attentive statistics pooling \cite{snyder2017}.
This observation suggests that frame-level attention can be unstable for this task, and applying attention at the segment level is essential for reliable performance.
Second, the results indicate a trade-off between ranking-based metrics and absolute error.
Global statistics pooling leads to the lowest prediction errors (MSE), which is also reflected in a higher linear correlation (LCC), while segmental attentive statistics pooling achieves the strongest rank correlations (SRCC, KTAU).
These respective strengths, which represent the best performance among all single-granularity baselines on different metrics, are underlined in Table \ref{tab:track1}.
This finding suggests that segment-level and global perspectives provide unique yet complementary insights into audio quality, a synergy that the DRASP framework is explicitly designed to leverage by integrating these two approaches.

Furthermore, we compared DRASP with pooling methods that similarly leverage multi-view concepts, such as multi-head \cite{india2019} and multi-resolution attentive pooling \cite{wang2020}.
Although these methods exhibit competitive performance, their architectures necessitate the computation of attention multiple times across the entire frame-level sequence, resulting in increased computational complexity.
By performing attention on a much shorter sequence of segments, DRASP provides a more efficient alternative while achieving superior results in our experiments.

\begin{figure}[t]
\centering
\includegraphics[width=1.0\linewidth]{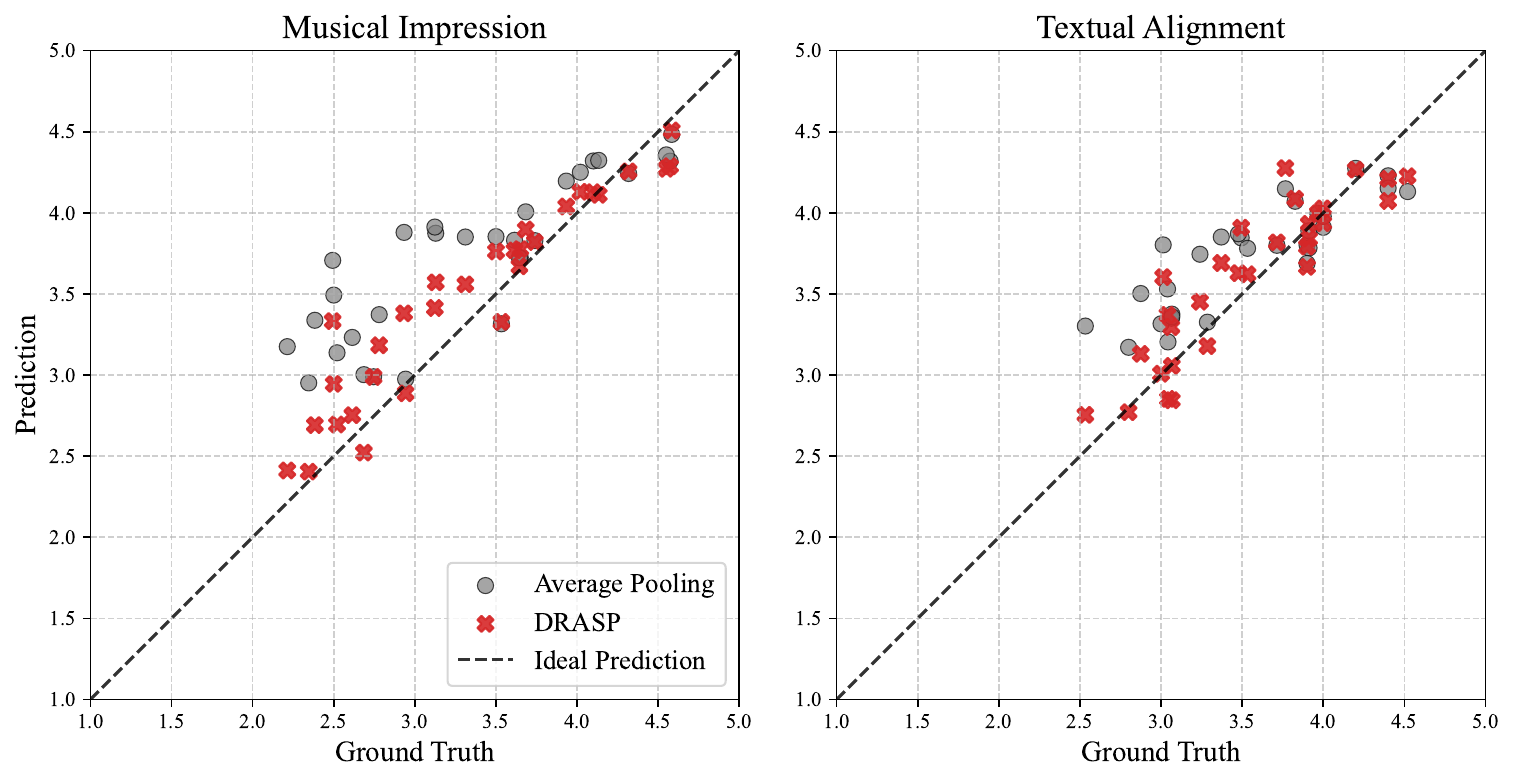}
\vspace{-20pt}
\caption{Scatter plots comparing the system-level performance of the proposed DRASP framework with a standard average pooling baseline on MusicEval.}
\label{fig:scatter}
\vspace{-15pt}
\end{figure}

\begin{figure}[t]
\centering
\includegraphics[width=1.0\linewidth]{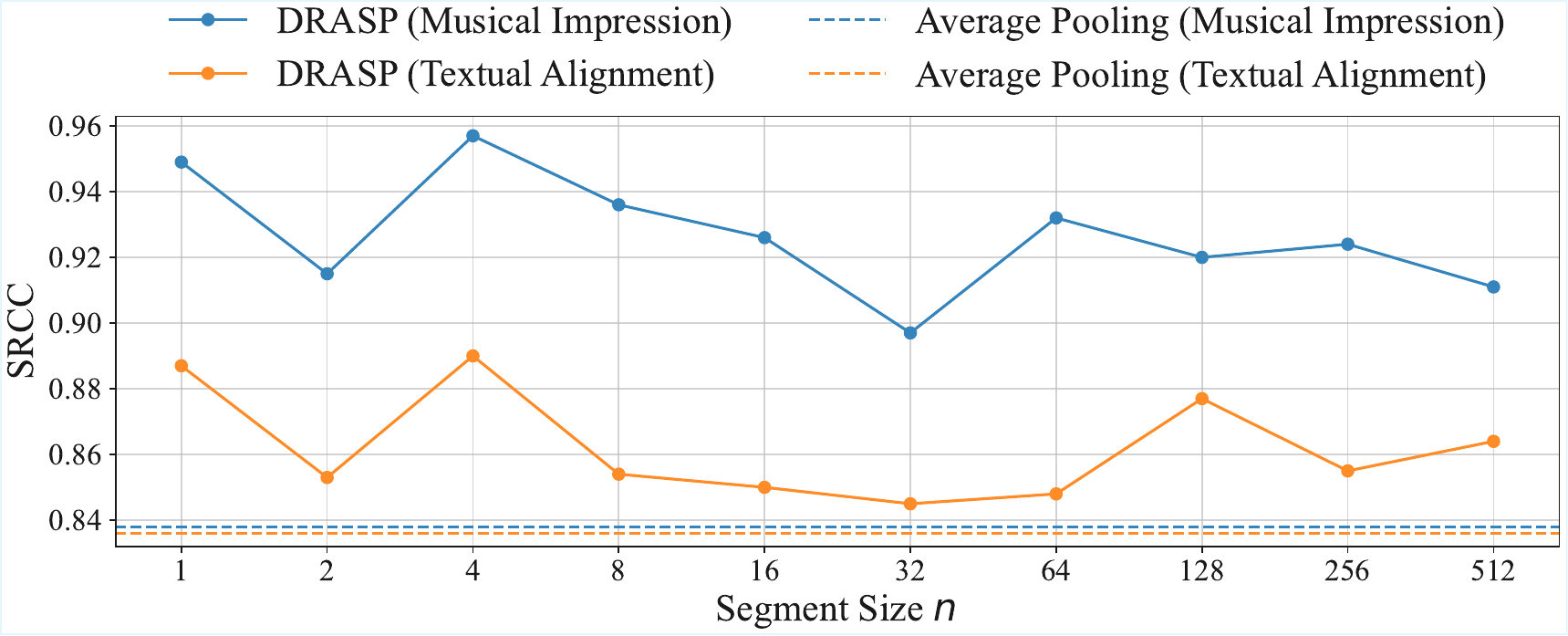}
\vspace{-20pt}
\caption{Effect of various segment sizes $n$ on system-level SRCC performance of our DRASP mechanism on MusicEval.}
\label{fig:segment_sizes}
\vspace{-5pt}
\end{figure}

\subsection{Visualization of Prediction Behavior}

To provide a qualitative understanding of the prediction behavior of our model, Fig. \ref{fig:scatter} presents scatter plots that map system-level predicted MOS scores against ground-truth human ratings.
The alignment of points along the dashed ideal prediction line provides an intuitive visual measure of performance, where a tighter and more linear clustering indicates lower prediction error and stronger correlations with human ratings.
As illustrated in both panels of Fig. \ref{fig:scatter}, the predictions derived from DRASP exhibit a considerably tighter and more linear cluster around the ideal line when contrasted with the dispersed results of the average pooling baseline.
This visualization transcends mere confirmation of the lower MSE outlined in Table \ref{tab:track1}.
It qualitatively demonstrates the superior stability and reliability of the predictions generated by DRASP.
The reduced variance suggests that our dual-resolution mechanism yields more consistent and well-calibrated assessments of audio quality across different generative systems.

\begin{table}[t]
\centering
\setlength{\tabcolsep}{4.5pt}
\caption{System-level SRCC results on the AES-Natural dataset for assessing model generalization.}
\vspace{-5pt}
\label{tab:track2}
\begin{tabular}{lccccc}
\toprule
\bf{Method} & \bf{PQ} & \bf{PC} & \bf{CE} & \bf{CU} & \bf{Avg.} \\
\toprule
AudioBox-Aesthetics & 0.877 & 0.935 & 0.839 & 0.832 & 0.871 \\
CLAP-based + DRASP & 0.881 & \bf{0.938} & 0.829 & 0.877 & 0.881 \\
AudioBox-Aesthetics + DRASP & \bf{0.900} & 0.936 & \bf{0.890} & \bf{0.911} & \bf{0.909} \\
\bottomrule
\end{tabular}
\vspace{-15pt}
\end{table}

\subsection{Effect of Various Segment Sizes}

Fig. \ref{fig:segment_sizes} presents the impact of segment size $n$ on the performance of the segment-wise average pooling module.
Our DRASP mechanism significantly outperforms the average pooling baseline across all evaluated segment sizes, underscoring the robustness of its underlying architecture.
Additionally, as shown in the plot, the selection of segment size markedly affects our model's performance, highlighting a crucial trade-off in temporal aggregation.
A smaller segment size allows the model to capture fine-grained acoustic details but may be overly sensitive to localized noise or minor imperfections.
Conversely, a larger segment size provides a smoother representation by averaging over more frames, but risks blurring distinct perceptual events and losing temporal precision.
This analysis reveals the role of the segment size as a controllable factor for balancing local fidelity with contextual stability.
Thus, the selection of an appropriate segment size is essential for optimizing the trade-off between sensitivity to detailed features and the richness of temporal context, directly affecting the overall efficacy of our model.

\subsection{Analysis of Generalization Capability on AES-Natural}

To assess the generalization capability and modularity of our DRASP mechanism, we conducted experiments on the diverse and challenging AES-Natural dataset.
As shown in Table \ref{tab:track2}, we compared three models: the strong AudioBox-Aesthetics model \cite{tjandra2025} pre-trained on over 500 hours of diverse audio, the CLAP-based model \cite{liu2025} enhanced with DRASP, and the same AudioBox-Aesthetics model also enhanced with DRASP.
We ensured a fair comparison by fine-tuning all models, including the original AudioBox-Aesthetics baseline, on the same in-domain training data before evaluation.
Notably, integrating DRASP into the AudioBox-Aesthetics backbone (AudioBox-Aesthetics + DRASP) yields the best overall performance, achieving the highest average SRCC and leading in three of the four evaluated dimensions.
These results indicate that DRASP functions as a versatile pooling module that generalizes across architectures and consistently enhances performance.

Beyond architectural compatibility, DRASP also demonstrates remarkable data efficiency.
Even when fine-tuned on only 35 hours of data, the DRASP-enhanced CLAP-based model (CLAP-based + DRASP) surpasses the AudioBox-Aesthetics model, which relies on pre-training with a substantially larger dataset.
This highlights the ability of DRASP to deliver competitive results under limited data conditions.
In summary, the consistent improvements across various backbones and data regimes reaffirm the robust generalization capability and practical applicability of the proposed DRASP as a modular pooling solution.

\section{Conclusion and Future Work}

This study introduces DRASP, an innovative pooling mechanism for automatic MOS prediction designed to address the shortcomings of conventional methods that uniformly treat all segments of an audio clip.
By combining a global statistical view with a segment-level attention mechanism that selectively emphasizes perceptually salient regions, DRASP captures both coarse and fine-grained information relevant to human judgment.
Extensive experiments conducted on the MusicEval and AES-Natural datasets reveal that DRASP consistently surpasses existing pooling methods, yielding reduced prediction errors and improved correlations with human ratings.

Looking ahead, a promising direction is the development of an adaptive segmentation strategy that empowers the model to learn meaningful boundaries directly from the data, such as those aligned with phonemes, musical notes, or distinct acoustic events.
Such a data-driven approach would facilitate more precise and perceptually aligned attention mechanisms.
Moreover, the current dual-resolution framework can be generalized into a multi-resolution architecture that simultaneously captures features across multiple temporal scales, including short and long segments.
This hierarchical modeling could further enhance the system's ability to capture complex temporal dynamics and improve robustness across diverse audio types.

\bibliographystyle{IEEEtran}
\bibliography{references}
\end{document}